\documentclass[12pt,a4paper]{article}
\usepackage{cite}
\usepackage{fullpage}
\usepackage[centertags]{amsmath}
\allowdisplaybreaks[4]
\usepackage{amssymb}
\usepackage{bm}
\usepackage[dvips]{graphicx}
\usepackage{url}
\usepackage[dvips,hyperindex]{hyperref}

\newcommand{\boss}[2]{\ensuremath{\rlap{\kern-2.5pt\ensuremath{\overset{\scriptscriptstyle(-)}{\phantom{#1}}}}{\ensuremath{{#1}_{#2}}}}}
\begin{document}

\begin{flushright}
\begin{tabular}{r}
\texttt{arXiv:0902.1992v2 [hep-ph]}
\\
\textsf{16 July 2009}
\end{tabular}
\end{flushright}
\vspace{1cm}
\begin{center}
\Large\bfseries
VSBL Electron Neutrino Disappearance
\\[0.5cm]
\large\normalfont
Carlo Giunti\ensuremath{^{(a)}}, Marco Laveder\ensuremath{^{(b)}}
\\[0.5cm]
\normalsize\itshape
\setlength{\tabcolsep}{1pt}
\begin{tabular}{cl}
\ensuremath{(a)}
&
INFN, Sezione di Torino,
Via P. Giuria 1, I--10125 Torino, Italy
\\[0.3cm]
\ensuremath{(b)}
&
Dipartimento di Fisica ``G. Galilei'', Universit\`a di Padova,
and
\\
&
INFN, Sezione di Padova,
Via F. Marzolo 8, I--35131 Padova, Italy
\end{tabular}
\end{center}
\begin{abstract}
We consider possible indications of
Very-Short-BaseLine (VSBL) electron neutrino disappearance
into sterile neutrinos
in MiniBooNE neutrino data and Gallium radioactive source experiments.
We discuss the compatibility of such a disappearance with reactor and MiniBooNE antineutrino data.
We find a tension between neutrino and antineutrino data which could be due to:
1) statistical fluctuations;
2) underestimate of systematic uncertainties;
3) exclusion of our hypothesis of VSBL $\nu_{e}$ disappearance;
4) a violation of CPT symmetry.
Considering the first possibility,
we present the results of a combined fit of all data, which indicate that
$ P_{\nu_{e}\to\nu_{e}} < 1 $
with 97.04\% CL.
We consider also the possibility of CPT violation,
which leads to the best-fit value
$ A_{ee}^{\text{CPT,bf}} = -0.17 $
for the asymmetry of the $\nu_{e}$ and $\bar\nu_{e}$ survival probabilities
and
$A_{ee}^{\text{CPT}}<0$ at 99.7\% CL.
\end{abstract}

\newpage

\section{Introduction}
\label{Introduction}
\nopagebreak

The MiniBooNE collaboration confirmed recently \cite{0812.2243}
the initial results on the search for short-baseline $\nu_{\mu}\to\nu_{e}$ oscillations
reported in Ref.~\cite{0704.1500}.
The new analysis confirms the absence of a signal in the $475-3000$ MeV energy range due to $\nu_{\mu}\to\nu_{e}$ oscillations
with a $\Delta{m}^2$ compatible with the indication of $\bar\nu_{\mu}\to\bar\nu_{e}$ oscillations
found in the LSND experiment \cite{hep-ex/0104049}.
It confirms also the anomalous excess of low-energy $\nu_{e}$-like events
reported in Ref.~\cite{0704.1500}.
The MiniBooNE collaboration have also presented recently
the first results for the $\bar\nu_{\mu}\to\bar\nu_{e}$ channel \cite{Karagiorgi-FNAL-2008,0904.1958},
which however are not as precise as the neutrino data
because of a much lower statistics.
Again, there is no evidence for a signal in the $475-3000$ MeV energy range due to $\bar\nu_{\mu}\to\bar\nu_{e}$ oscillations
compatible with the indication found in the LSND experiment.
More interesting,
there is also no evidence of a low-energy anomalous excess of events.

In Ref.~\cite{0707.4593} we proposed an explanation\footnote{
Other explanations have been proposed in Refs.~\cite{0708.1281,0712.1230,0709.4004,0711.1363}.
}
of the MiniBooNE low-energy anomaly
through the Very-Short-BaseLine (VSBL) disappearance of $\nu_{e}$'s due to oscillations into sterile neutrinos
generated by a large $\Delta{m}^2$ in the range
\begin{equation}
20 \, \text{eV}^{2}
\lesssim
\Delta{m}^{2}
\lesssim
330 \, \text{eV}^{2}
\,,
\label{dm2}
\end{equation}
which is motivated \cite{Laveder:2007zz,hep-ph/0610352,0707.4593}
by the anomalous ratio of measured and predicted ${}^{71}\text{Ge}$
production rates,
\begin{equation}
R_{\text{Ga}}
=
0.87 \pm 0.05
\qquad
\text{\protect\cite{0901.2200}}
\,,
\label{ga}
\end{equation}
observed in the Gallium radioactive source experiments
GALLEX
\cite{Anselmann:1995ar,Hampel:1998fc}
and SAGE
\cite{Abdurashitov:1996dp,hep-ph/9803418,nucl-ex/0512041,0901.2200}.
As shown in Ref.~\cite{0707.4593},
the range of $\Delta{m}^{2}$ in Eq.~(\ref{dm2})
is compatible with the upper bounds on neutrino masses
obtained in tritium $\beta$-decay and neutrinoless double-$\beta$ decay experiments.

The large $\Delta{m}^{2}$ in Eq.~(\ref{dm2}) implies
an oscillation length of MiniBooNE neutrinos and antineutrinos
which is shorter than the source-detector distance
of about 541 m:
\begin{equation}
L_{\text{osc}}^{\text{MB}}
=
\frac{ 4 \pi E }{ \Delta{m}^{2} }
\lesssim
400 \, \text{m}
\,,
\label{Losc}
\end{equation}
where $E<3\,\text{GeV}$ is the neutrino energy.
Therefore,
the disappearance probability of electron neutrinos and antineutrinos
averaged in each energy bin
is approximately constant\footnote{
The analysis of MiniBooNE data with an energy-dependent $\nu_{e}$ disappearance probability
due to a $ \Delta{m}^{2} \lesssim 20 \, \text{eV}^{2} $
and the compatibility with Gallium and reactor neutrino data \protect\cite{0711.4222}
will be discussed elsewhere
\protect\cite{Giunti-Laveder-IP-09}.
}
over the MiniBooNE energy spectrum
(CPT implies that $P_{\nu_{e}\to\nu_{e}}=P_{\bar\nu_{e}\to\bar\nu_{e}}$;
see Ref.~\cite{Giunti-Kim-2007}).

At first sight, explaining the anomalous MiniBooNE low-energy excess of $\nu_{e}$-like events with
$\nu_{e}$ disappearance seems a contradiction.
Indeed,
we must introduce another ingredient,
a factor $f_{\nu}$ which takes into account the uncertainty
of the normalization of the background prediction,
which receives a substantial contribution from the uncertainty
of the normalization of the 
calculated neutrino flux (see Ref.\cite{physics/0609129}).
The background events in the MiniBooNE data analysis are divided in
$\nu_{e}$-induced events
and
misidentified $\nu_{\mu}$-induced events.
The $\nu_{e}$-induced events are produced by the $\nu_{e}$'s in the beam generated at the source
by pion and kaon decays.
The number of $\nu_{e}$-induced events ($N_{\nu_{e}}^{\text{cal}}$)
is smaller than the number of misidentified $\nu_{\mu}$-induced events ($N_{\nu_{\mu}}^{\text{cal}}$)
in the low-energy bins
and larger in the high-energy bins
(see Fig.~1 of Ref.~\cite{0812.2243}).
An increase of the total background by a factor $f_{\nu}$ leads to a fit of the low-energy excess through
the increase of the dominant misidentified $\nu_{\mu}$-induced events.
The high-energy bins are fitted by compensating the increase of the dominant $\nu_{e}$-induced events
by the factor $f_{\nu}$ with the disappearance of $\nu_{e}$'s.

A normalization factor $f_{\nu}$ significantly different from unity
is allowed by the 15\% uncertainty of the calculated neutrino flux \cite{0806.1449}.
This uncertainty is consistent with the
measured ratio $ 1.21 \pm 0.24 $
of detected and predicted charged-current quasi-elastic $\nu_{\mu}$ events \cite{0706.0926}.
Although the background calculated by the MiniBooNE collaboration
has been normalized to such measured number of charged-current quasi-elastic
$\nu_{\mu}$ events,
an uncertainty of about 15\% remains.
We consider this value as a reliable
estimate of the uncertainty of the background prediction,
which is more conservative than that estimated in Ref.~\cite{0812.2243}.

In this paper we update the analysis presented in Ref.~\cite{0707.4593}
of MiniBooNE neutrino events by
considering the new data in Ref.~\cite{0812.2243} (Section~\ref{neu}).
We discuss the compatibility of MiniBooNE neutrino data
with Gallium (Section~\ref{neu-gal}),
reactor (Section~\ref{neu-gal-rea})
and
MiniBooNE antineutrino (Section~\ref{anu})
data.
We also consider the possibility of a violation
of the CPT equality
$ P_{\nu_{e}\to\nu_{e}} = P_{\bar\nu_{e}\to\bar\nu_{e}} $
(Section~\ref{CPT}).

\section{MiniBooNE Neutrino Data}
\label{neu}
\nopagebreak

We consider the MiniBooNE $\nu_{e}$-like events from Ref.~\cite{0812.2243,Louis-private-09},
which are listed in Tab.~\ref{tab-neu-data}.
We fit these data with the theoretical hypothesis
\begin{equation}
N^{\text{the}}_{\nu,j}
=
f_{\nu} \left(
P_{\nu_{e}\to\nu_{e}} N_{\nu_{e},j}^{\text{cal}} + N_{\nu_{\mu},j}^{\text{cal}}
\right)
\,,
\label{teo-neu}
\end{equation}
where $N_{\nu_{e},j}^{\text{cal}}$ and $N_{\nu_{\mu},j}^{\text{cal}}$
are, respectively,
the calculated number of expected $\nu_{e}$-induced and misidentified $\nu_{\mu}$-induced events
in the third and fourth columns of Tab.~\ref{tab-neu-data}.
We calculate the best fit values of the parameters
$P_{\nu_{e}\to\nu_{e}}$ and $f_{\nu}$ by minimizing the least-square function
\begin{equation}
\chi^2_{\text{MB-$\nu$}}
=
\sum_{j=1}^{11}
\frac
{ \left( N^{\text{the}}_{\nu,j} - N^{\text{exp}}_{\nu,j} \right)^2 }
{ N^{\text{the}}_{\nu,j} }
+
\left( \frac{ f_{\nu} - 1 }{ \Delta f_{\nu} } \right)^2
\,,
\label{chi-mib-neu}
\end{equation}
with \cite{0806.1449}
\begin{equation}
\Delta f_{\nu} = 0.15
\,,
\label{dfnu}
\end{equation}
according to the discussion in the introductory Section~\ref{Introduction}.

\begin{figure}[p!]
\begin{center}
\includegraphics*[bb=35 147 562 695, width=0.48\textwidth]{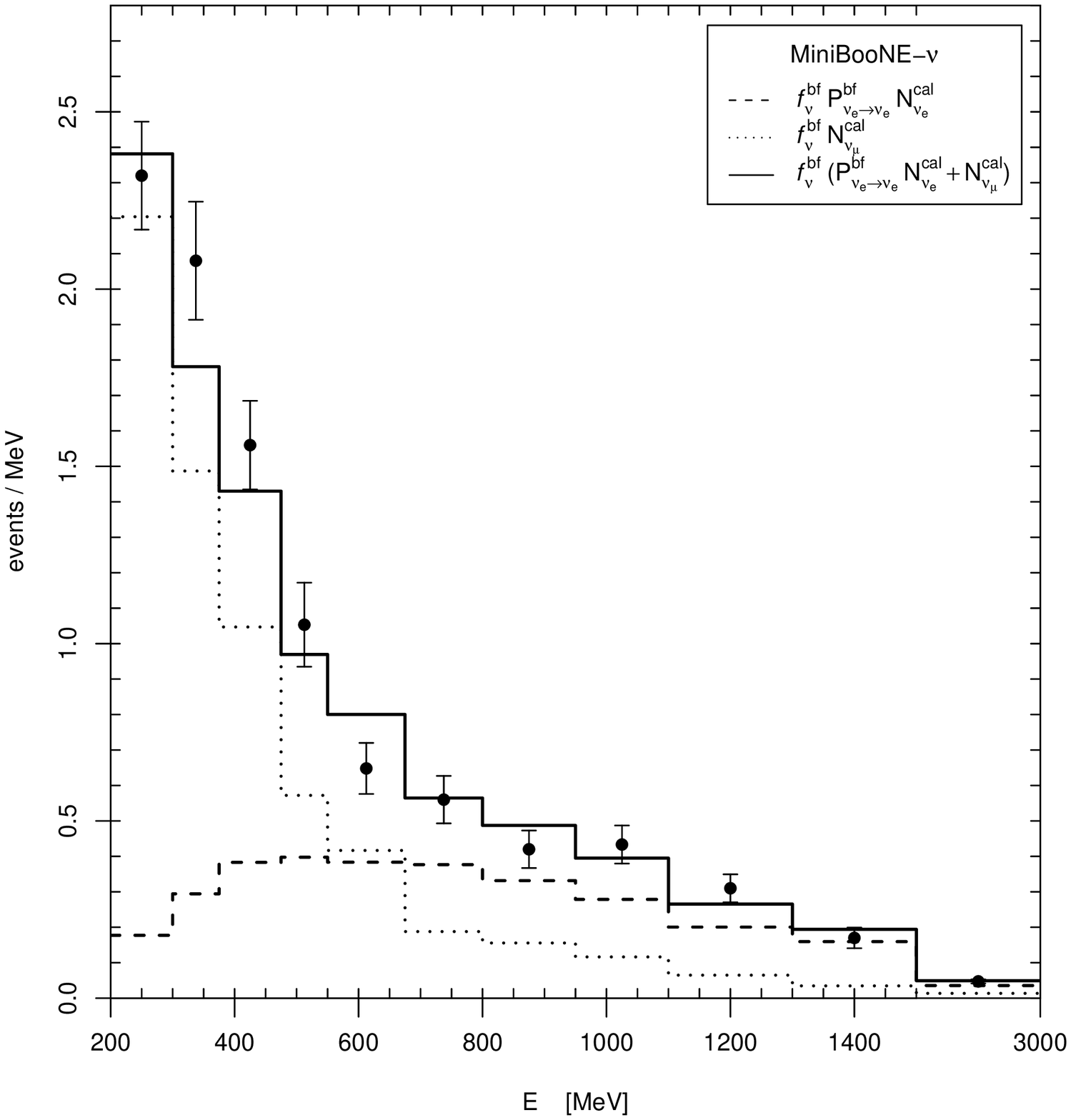}
\caption{ \label{hst-neu}
Theoretically expected number of $\nu_{e}$ events compared with MiniBooNE data,
represented by the points with their statistical error bars.
The values of $f_{\nu}^{\text{bf}}$ and $P_{\nu_{e}\to\nu_{e}}^{\text{bf}}$
are those in Tab.~\ref{tab-bef-neu},
corresponding to the best fit of MiniBooNE neutrino data
(MB-$\nu$).
}
\end{center}
\end{figure}

The results of the minimization of $\chi^2_{\text{MB-$\nu$}}$ are presented in Tab.~\ref{tab-bef-neu}
in the MB-$\nu$ column.
One can see that the hypothesis of $\nu_{e}$ disappearance improves the fit 
with a significant decrease of $\chi^2_{\text{MB-$\nu$}}$ from
27.2 to 17.7 .
Note that in both cases we minimize the $\chi^2$ with respect to $f_{\nu}$,
which is an intrinsic uncertainty in the MiniBooNE result.
As shown in Fig.~\ref{hst-neu},
the rather large best-fit value $f_{\nu}^{\text{bf}}=1.31 $,
allows us to fit the low-energy excess by increasing the
number of misidentified $\nu_{\mu}$-induced events,
whereas the rather low best-fit value $P_{\nu_{e}\to\nu_{e}}^{\text{bf}}=0.72 $
maintains a good fit of the intermediate-energy and high-energy data
by avoiding an increase of the $\nu_{e}$-induced events through the product
$ f_{\nu}^{\text{bf}} P_{\nu_{e}\to\nu_{e}}^{\text{bf}} = 0.94 $.

\begin{figure}[p!]
\begin{center}
\includegraphics*[bb=30 147 572 704, width=0.48\textwidth]{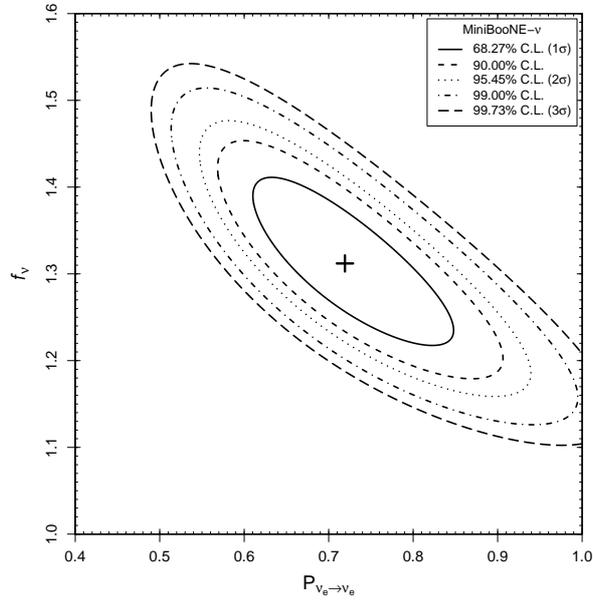}
\caption{ \label{chi-neu-con}
Allowed regions in the
$P_{\nu_{e}\to\nu_{e}}$--$f_{\nu}$
plane
obtained from the fit of MiniBooNE neutrino data.
}
\end{center}
\end{figure}

Figure~\ref{chi-neu-con} shows the allowed regions in the
$P_{\nu_{e}\to\nu_{e}}$--$f_{\nu}$
plane.
One can see that the uncertainties for the two parameters are correlated.
The reason is that the product
$ f_{\nu} P_{\nu_{e}\to\nu_{e}} $
is constrained by the fitting of the intermediate-energy and high-energy data
through the dominating $\nu_{e}$-induced events.

The $\Delta\chi^2=\chi^2_{\text{MB-$\nu$}}-\chi^2_{\text{MB-$\nu$,min}}$
marginalized over $f_{\nu}$ is shown in Fig~\ref{chi-plt}
as a function of $P_{\nu_{e}\to\nu_{e}}$.
One can see that the MiniBooNE neutrino data indicate $ P_{\nu_{e}\to\nu_{e}} < 1 $
with 99.80\% CL.
The allowed ranges of $P_{\nu_{e}\to\nu_{e}}$ at different values of confidence level
are listed in Tab.~\ref{tab-pee}.

\section{Gallium Radioactive Source Experiments}
\label{neu-gal}
\nopagebreak

In this Section we discuss the results of a combined fit of
the MiniBooNE neutrino data considered in the previous Section with the result
in Eq.~(\ref{ga}) for the ratio of measured and predicted ${}^{71}\text{Ge}$
production rates
in the Gallium radioactive source experiments \cite{Anselmann:1995ar,Hampel:1998fc,Abdurashitov:1996dp,hep-ph/9803418,nucl-ex/0512041,0901.2200}.

We consider the sum of the MiniBooNE neutrino least-square function in Eq.~(\ref{chi-mib-neu})
with the Gallium least-square function
\begin{equation}
\chi^2_{\text{Ga}}
=
\left(
\frac{ P_{\nu_{e}\to\nu_{e}} - R_{\text{Ga}} }{ \Delta R_{\text{Ga}} }
\right)^2
\,,
\label{chi-gal}
\end{equation}
where, from Eq.~(\ref{ga}), $R_{\text{Ga}}=0.87$ and $\Delta R_{\text{Ga}}=0.05$.
The results of the minimization of
$\chi^2=\chi^2_{\text{MB-$\nu$}}+\chi^2_{\text{Ga}}$
are listed in Tab.~\ref{tab-bef-neu}
in the MB-$\nu$+Ga column.
The parameter goodness-of-fit \cite{hep-ph/0304176} of 12.4\% 
implies that the results of the MiniBooNE neutrino and
the Gallium radioactive source experiments are compatible
in the framework of the VSBL $\nu_{e}$ disappearance hypothesis.
The best-fit values
$f_{\nu}^{\text{bf}}=1.24 $
and
$P_{\nu_{e}\to\nu_{e}}^{\text{bf}}=0.83 $,
are more reasonable than those obtained from the fit of MiniBooNE neutrino data alone.
The goodness of fit of 2.8\% is acceptable and
much better than the 0.04\% obtained without $\nu_{e}$ disappearance.
Figure~\ref{chi-plt} shows
$\Delta\chi^2=\chi^2-\chi^2_{\text{min}}$
marginalized over $f_{\nu}$
as a function of $P_{\nu_{e}\to\nu_{e}}$.
One can see that $P_{\nu_{e}\to\nu_{e}}=1$ is disfavored at more than $3\sigma$
(the precise value is 99.98\% CL).
The allowed ranges of $P_{\nu_{e}\to\nu_{e}}$ at different values of confidence level
are listed in Tab.~\ref{tab-pee}.

As a caveat,
we need to mention that another possible explanation of the Gallium anomaly\footnote{
The authors of Ref.~\cite{0805.2098} proposed yet another explanation.
}
is that
the theoretical
cross section of the Gallium detection process in Ref.~\cite{hep-ph/9710491},
which has been used in deriving Eq.~(\ref{ga}),
has been overestimated \cite{nucl-ex/0512041,hep-ph/0605186,0901.2200}.
This is possible, because the detection process
$ \nu_{e} + {}^{71}\text{Ga} \to {}^{71}\text{Ge} + e^{-} $
can populate excited states of ${}^{71}\text{Ge}$
with transition amplitudes which have large uncertainties
\cite{nucl-th/9503017,nucl-th/9804011}.
Let us however remark that shell model calculations indicate \cite{nucl-th/9804011} that
the cross section of the Gallium detection process could be even larger than that used in deriving Eq.~(\ref{ga}),
leading to a stronger anomaly.

\section{Reactor Neutrino Experiments}
\label{neu-gal-rea}
\nopagebreak

The indication in favor of VSBL $\nu_{e}$ disappearance that we have obtained from the data of
the MiniBooNE neutrino experiment and
the Gallium radioactive source experiments
must be confronted with the lack of any evidence of VSBL $\bar\nu_{e}$ disappearance
in reactor experiments\footnote{
The compatibility of MiniBooNE neutrino data with the weak indication of
$\bar\nu_{e}$ disappearance due $ \Delta{m}^{2} \simeq 2 \, \text{eV}^{2} $
found in Ref.~\cite{0711.4222}
from the analysis of the Bugey reactor neutrino data \cite{Declais:1995su}
will be discussed elsewhere
\protect\cite{Giunti-Laveder-IP-09}.
},
since
$P_{\nu_{e}\to\nu_{e}}=P_{\bar\nu_{e}\to\bar\nu_{e}}$
in the framework of local CPT-invariant Quantum Field Theory
(see Ref.~\cite{Giunti-Kim-2007}).

Here we will consider the data of the following reactor experiments
($R^{\text{(d)}}$ denotes the ratio of measured and predicted event rates at the source-detector distance $d$):

\begin{description}

\item[Gosgen] \cite{Zacek:1986cu}:
\begin{align}
\null & \null
R_{\text{Gosgen}}^{\text{(37.9\,m)}}
=
1.018 \pm 0.019 \pm 0.015 \pm 0.060
\,,
\label{gosgen-1}
\\
\null & \null
R_{\text{Gosgen}}^{\text{(45.9\,m)}}
=
1.045 \pm 0.019 \pm 0.015 \pm 0.060
\,,
\label{gosgen-2}
\\
\null & \null
R_{\text{Gosgen}}^{\text{(64.7\,m)}}
=
0.975 \pm 0.036 \pm 0.030 \pm 0.060
\,,
\label{gosgen-3}
\end{align}
where the first uncertainty is statistic,
the second is uncorrelated systematic and the third is correlated systematic.

\item[Bugey] \cite{Declais:1995su}:
\begin{align}
\null & \null
R_{\text{Bugey}}^{\text{(15\,m)}}
=
0.988 \pm 0.004 \pm 0.05
\,,
\label{bugey-1}
\\
\null & \null
R_{\text{Bugey}}^{\text{(40\,m)}}
=
0.994 \pm 0.010 \pm 0.05
\,,
\label{bugey-2}
\\
\null & \null
R_{\text{Bugey}}^{\text{(95\,m)}}
=
0.915 \pm 0.132 \pm 0.05
\,.
\label{bugey-3}
\end{align}

\item[Chooz] \cite{hep-ex/0301017}:
\begin{equation}
R_{\text{Chooz}}^{\text{(1\,km)}}
=
1.01 \pm 0.028 \pm 0.027
\,.
\label{chooz}
\end{equation}

\end{description}

The Gosgen, Bugey and Chooz collaborations estimated, respectively,
a 3.0\% \cite{Zacek:1986cu},
a 2.8\% \cite{Declais:1995su}, and
a 1.9\% \cite{hep-ex/0301017}
systematic uncertainty of the reactor neutrino flux.
The estimate of the Chooz collaboration was based on the assumption that previous reactor experiments
measured the reactor neutrino flux without any disappearance.
Since we are considering the possibility of such a disappearance,
we must increase the Chooz systematic uncertainty of the reactor neutrino flux
to the same value as that of Gosgen and Bugey, i.e. approximately 3\%,
leading to
\begin{equation}
R_{\text{Chooz}}^{\text{(1\,km)}}
=
1.01 \pm 0.028 \pm 0.036
\,.
\label{chooz-cor}
\end{equation}
Taking a 3\% systematic uncertainty of the reactor neutrino flux correlated in all reactor neutrino experiments,
we obtain the covariance matrix of reactor data
\begin{equation}
V_{\text{Re}}
=
\begin{pmatrix}
4.19 & 3.60 & 3.60 & 0.90 & 0.90 & 0.90 & 0.90\\
3.60 & 4.19 & 3.60 & 0.90 & 0.90 & 0.90 & 0.90\\
3.60 & 3.60 & 5.80 & 0.90 & 0.90 & 0.90 & 0.90\\
0.90 & 0.90 & 0.90 & 2.63 & 2.62 & 2.62 & 0.90\\
0.90 & 0.90 & 0.90 & 2.62 & 2.72 & 2.62 & 0.90\\
0.90 & 0.90 & 0.90 & 2.62 & 2.62 & 20.04 & 0.90\\
0.90 & 0.90 & 0.90 & 0.90 & 0.90 & 0.90 & 2.05\\
\end{pmatrix}
\times
10^{-3}
\,,
\label{Vrea}
\end{equation}
for the vector of data
\begin{equation}
\vec{R}_{\text{Re}}
=
\left(
R_{\text{Gosgen}}^{\text{(37.9\,m)}},
R_{\text{Gosgen}}^{\text{(45.9\,m)}},
R_{\text{Gosgen}}^{\text{(64.7\,m)}},
R_{\text{Bugey}}^{\text{(15\,m)}},
R_{\text{Bugey}}^{\text{(40\,m)}},
R_{\text{Bugey}}^{\text{(95\,m)}},
R_{\text{Chooz}}^{\text{(1\,km)}}
\right)
\,.
\label{vecRe}
\end{equation}
We fit the reactor neutrino data by minimizing the leas-squares function
\begin{equation}
\chi^2_{\text{Re}}
=
\left( P_{\nu_{e}\to\nu_{e}} - \vec{R}_{\text{Re}} \right)^{T}
V_{\text{Re}}^{-1}
\left( P_{\nu_{e}\to\nu_{e}} - \vec{R}_{\text{Re}} \right)
\,.
\label{chi-rea}
\end{equation}
From the results presented in
Tab.~\ref{tab-bef-neu}
and
Fig.~\ref{chi-plt}
one can see that reactor data do not show any indication of $\bar\nu_{e}$ disappearance.

However,
the lower limits for
$ P_{\nu_{e}\to\nu_{e}} $
that one can infer from Fig.~\ref{chi-plt}
(see Tab.~\ref{tab-pee})
indicate that reactor data allow a small $\bar\nu_{e}$ disappearance.
Therefore, we tried a combined analysis of
MiniBooNE neutrino, Gallium and reactor data
under the hypothesis of $\nu_{e}$ disappearance
with
$P_{\nu_{e}\to\nu_{e}}=P_{\bar\nu_{e}\to\bar\nu_{e}}$.

The results of the minimization of
$ \chi^2 = \chi^2_{\text{MB-$\nu$}} + \chi^2_{\text{Ga}} + \chi^2_{\text{Re}} $
are given in the corresponding column in Tab.~\ref{tab-bef-neu}.
The rather low parameter goodness-of-fit, 0.4\% ,
shows that there is tension between MiniBooNE and Gallium neutrino data on one side
and reactor antineutrino data on the other side.
The tension is mainly due to a conflict between MiniBooNE and reactor data,
which have a
0.3\% 
parameter goodness-of-fit
($ \Delta\chi^{2}_{\text{min}} = 8.6 $
with
$ \text{NDF} = 1 $),
whereas
Gallium and reactor data have a
3.1\% 
parameter goodness-of-fit
($ \Delta\chi^{2}_{\text{min}} = 4.6 $
with
$ \text{NDF} = 1 $).

Possible explanations of this tension could be:

\begin{enumerate}

\item
Statistical fluctuations.

\item
Systematic uncertainties have been underestimated.

\item
Our hypothesis of VSBL $\nu_{e}$ disappearance is excluded.

\item
There is a violation of CPT symmetry
leading to
$ P_{\nu_{e}\to\nu_{e}} \neq P_{\bar\nu_{e}\to\bar\nu_{e}} $.

\end{enumerate}

Considering the first possibility,
the combined fit of
MiniBooNE neutrino, Gallium and reactor data
leads to the
$\Delta\chi^2=\chi^2-\chi^2_{\text{min}}$
depicted in Fig.~\ref{chi-plt} and
the allowed ranges of $P_{\nu_{e}\to\nu_{e}}$ listed in Tab.~\ref{tab-pee}.
One can see that the addition of the reactor constraint
narrows the allowed range of $ P_{\nu_{e}\to\nu_{e}} $
and shifts it towards unity.
There is an indication that
$ P_{\nu_{e}\to\nu_{e}} < 1 $
with 97.74\% CL.

Figure~\ref{hst-neu-gal-rea} shows that the fit of the MiniBooNE neutrino data with the
best-fit values
$f_{\nu}^{\text{bf}}=1.19 $
and
$P_{\nu_{e}\to\nu_{e}}^{\text{bf}}=0.93 $
is not as good as that in Fig.~\ref{hst-neu} corresponding to the fit of
MiniBooNE neutrino data alone,
but it is acceptable.
We notice that the value of 
$f_{\nu}^{\text{bf}}$
is not much different from that obtained from the fit of
MiniBooNE neutrino data without oscillations
($f_{\nu}^{\text{bf}}=1.15 $; see Tab.~\ref{tab-bef-neu}).
It is only slightly larger, in order to keep the product
$ f_{\nu}^{\text{bf}} P_{\nu_{e}\to\nu_{e}}^{\text{bf}} = 1.11 $
close to unity for a good fit of the intermediate-energy and high-energy MiniBooNE neutrino data.

\section{MiniBooNE Antineutrino Data}
\label{anu}
\nopagebreak

The MiniBooNE collaboration presented recently
the first results on the search for $\bar\nu_{\mu}\to\bar\nu_{e}$ oscillations \cite{Karagiorgi-FNAL-2008,0904.1958}.
These data, as the neutrino data,
do not show any
evidence for a signal in the $475-3000$ MeV energy range due to $\bar\nu_{\mu}\to\bar\nu_{e}$ oscillations
compatible with the indication found in the LSND experiment.
There is also no evidence of a low-energy anomalous excess of events
analogous to that observed in the neutrino data.
However,
since in MiniBooNE the antineutrino statistics is about five times smaller than the neutrino statistics,
the uncertainties of the antineutrino data do not allow us to infer strong constraints on new physics.

\begin{figure}[p!]
\begin{center}
\includegraphics*[bb=24 147 572 702, width=0.48\textwidth]{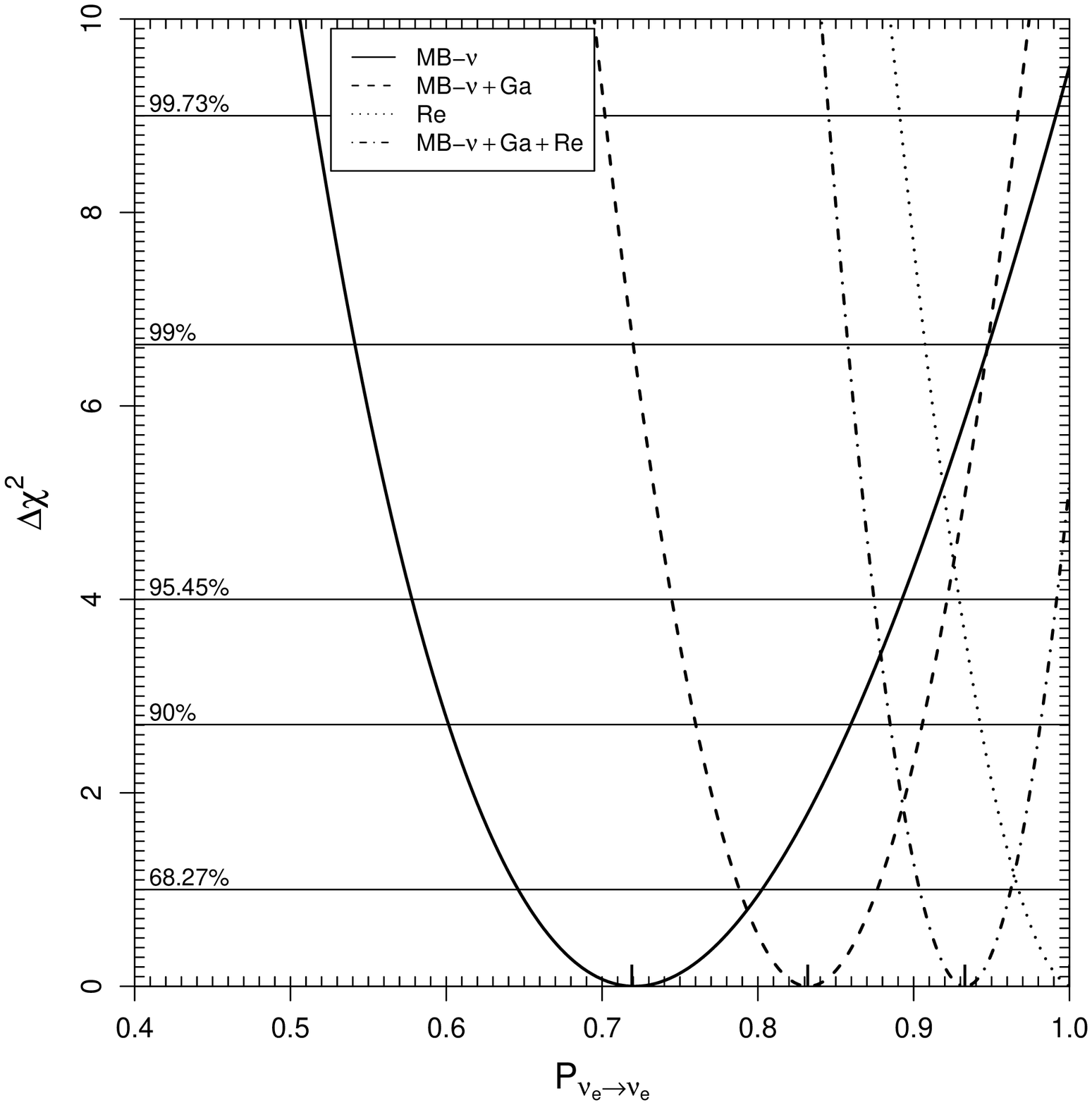}
\hfill
\includegraphics*[bb=24 147 572 702, width=0.48\textwidth]{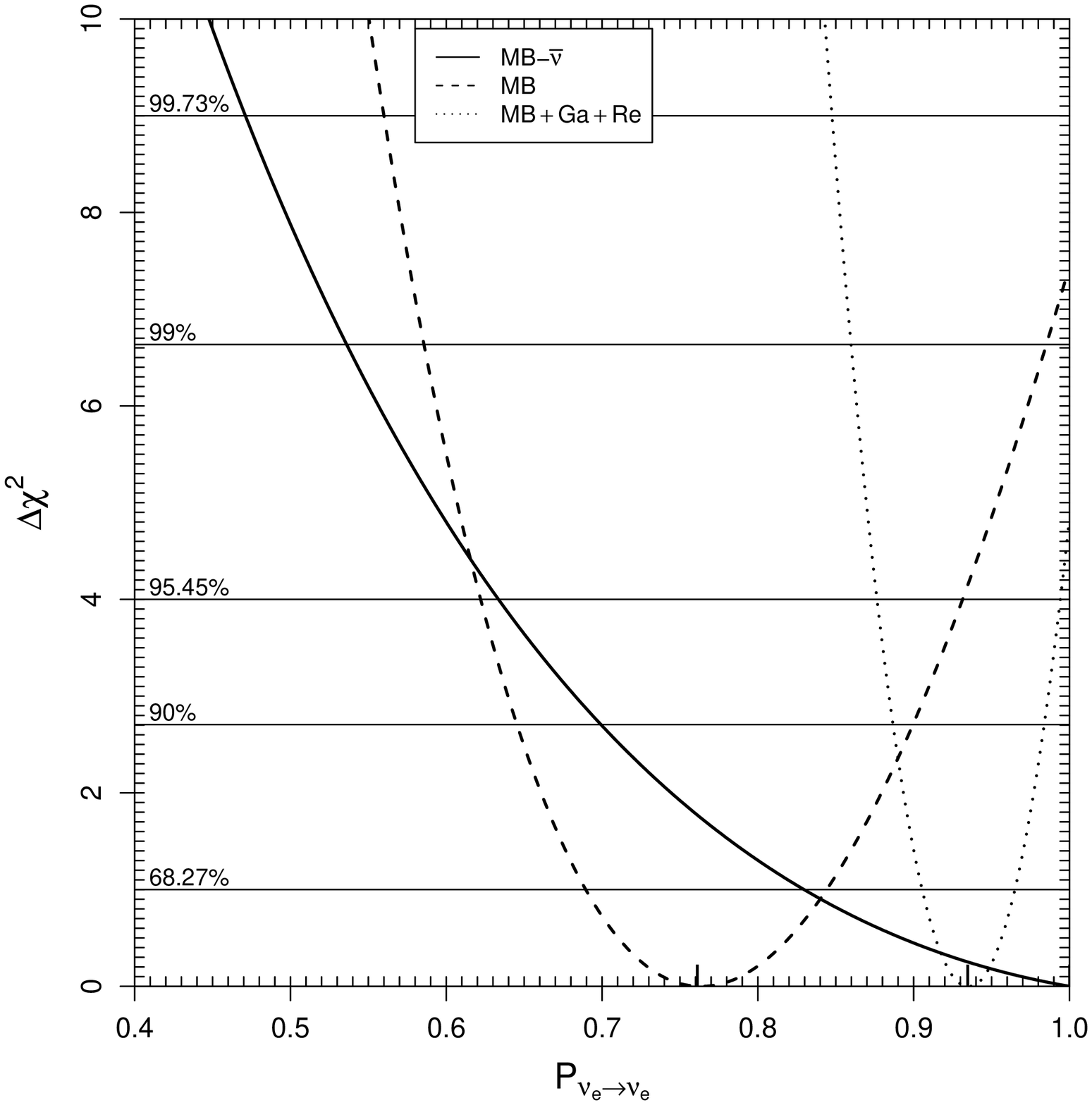}
\caption{ \label{chi-plt}
Marginal $\Delta\chi^2$'s
as a function of
$P_{\nu_{e}\to\nu_{e}}$
obtained from the fit of different combinations of
MiniBooNE neutrino (MB-$\nu$) and antineutrino (MB-$\bar\nu$), Gallium (Ga) and
reactor (Re) data.
The horizontal lines correspond to the indicated value of confidence level.
}
\end{center}
\end{figure}

\begin{figure}[p!]
\begin{center}
\includegraphics*[bb=35 146 562 695, width=0.48\textwidth]{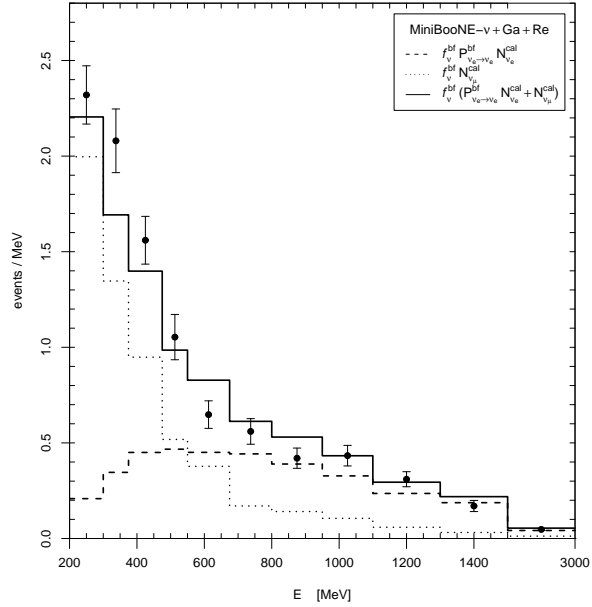}
\caption{ \label{hst-neu-gal-rea}
Theoretically expected number of $\nu_{e}$ events compared with MiniBooNE data,
represented by the points with their statistical error bars.
The values of $f_{\nu}^{\text{bf}}$ and $P_{\nu_{e}\to\nu_{e}}^{\text{bf}}$
are those in Tab.~\ref{tab-bef-neu},
corresponding to the best fit of MiniBooNE neutrino data, Gallium data
and reactor data
(MB-$\nu$+Ga+Re).
}
\end{center}
\end{figure}

In this section we study the compatibility of the 
VSBL $\nu_{e}$ disappearance hypothesis
with the MiniBooNE antineutrino data,
assuming the CPT equality $P_{\nu_{e}\to\nu_{e}}=P_{\bar\nu_{e}\to\bar\nu_{e}}$.
We fit the MiniBooNE antineutrino data with the theoretical hypothesis
\begin{equation}
N^{\text{the}}_{\bar\nu,j}
=
f_{\bar\nu} \left(
P_{\nu_{e}\to\nu_{e}} N_{\bar\nu_{e},j}^{\text{cal}} + N_{\bar\nu_{\mu},j}^{\text{cal}}
\right)
\,,
\label{teo-anu}
\end{equation}
where $N_{\bar\nu_{e},j}^{\text{cal}}$ and $N_{\bar\nu_{\mu},j}^{\text{cal}}$
are, respectively,
the calculated number of expected $\bar\nu_{e}$-induced and misidentified $\bar\nu_{\mu}$-induced events
in the third and fourth columns of Tab.~\ref{tab-anu-data}.
The least-squares function is
\begin{equation}
\chi^2_{\text{MB-$\bar\nu$}}
=
\sum_{j=1}^{11}
\frac
{ \left( N^{\text{the}}_{\bar\nu,j} - N^{\text{exp}}_{\bar\nu,j} \right)^2 }
{ N^{\text{the}}_{\bar\nu,j} }
+
\left( \frac{ f_{\bar\nu} - 1 }{ \Delta f_{\nu} } \right)^2
\,,
\label{chi-mib-anu}
\end{equation}
with \cite{0806.1449}
\begin{equation}
\Delta f_{\bar\nu} = 0.17
\,.
\label{dfan}
\end{equation}
The minimum of $\chi^2_{\text{MB-$\bar\nu$}}$ is obtained for
$P_{\nu_{e}\to\nu_{e}}=1$,
as shown in Tab.~\ref{tab-bef-anu}.
However,
the low statistics of the data do not allow us to
put stringent constraints on $P_{\nu_{e}\to\nu_{e}}$,
as one can see from the $\Delta\chi^2=\chi^2_{\text{MB-$\bar\nu$}}-\chi^2_{\text{MB-$\bar\nu$,min}}$
marginalized over $f_{\bar\nu}$ depicted in Fig.~\ref{chi-plt} (MB-$\bar\nu$ curve)
and the allowed ranges of $P_{\nu_{e}\to\nu_{e}}$ listed in Tab.~\ref{tab-pee}.

In fact, the parameter goodness-of-fit of 14.8\% 
of the combined fit of MiniBooNE neutrino and antineutrino data is acceptable.
The $\Delta\chi^2=\chi^2-\chi^2_{\text{min}}$,
where
$\chi^2=\chi^2_{\text{MB-$\nu$}}+\chi^2_{\text{MB-$\bar\nu$}}$,
depicted in Fig.~\ref{chi-plt} (MB curve)
shows that the MiniBooNE neutrino data are dominating.

Finally,
in Tab.~\ref{tab-bef-anu} and Fig.~\ref{chi-plt}
we present the results of the combined fit MB+Ga+Re of all the data considered so far:
the MiniBooNE neutrino and antineutrino data and
the Gallium and reactor data.
The parameter goodness-of-fit of 4.1\% 
do not allow us to reject the compatibility of the data
under the hypothesis of VSBL $\nu_{e}$ disappearance.
This results indicate that the possibility
that the tension between MiniBooNE neutrino and Gallium data on one side
and reactor data on the other side
is due to statistical fluctuations may be correct.
In this case, from Fig.~\ref{chi-plt} one can see that
$ P_{\nu_{e}\to\nu_{e}} < 1 $
with 97.04\% CL.
Therefore,
adding the MiniBooNE antineutrino data to the combined fit of
MiniBooNE, Gallium and reactor data does not change significantly the confidence level of the indication of
$ P_{\nu_{e}\to\nu_{e}} < 1 $
found in Section~\ref{neu-gal-rea}.
Indeed, the best fit values of $P_{\nu_{e}\to\nu_{e}}$ and $f_{\nu}$
(see Tabs.~\ref{tab-bef-neu} and \ref{tab-bef-anu})
and the allowed ranges of $P_{\nu_{e}\to\nu_{e}}$ listed in Tab.~\ref{tab-pee}
are practically the same as those obtained
without the MiniBooNE antineutrino data.
Therefore,
Fig.~\ref{hst-neu-gal-rea} and the related discussion at the end of Section~\ref{neu-gal-rea}
remain valid after the addition of MiniBooNE antineutrino data.

\section{CPT Violation}
\label{CPT}
\nopagebreak

\begin{figure}[p!]
\begin{center}
\includegraphics*[bb=24 147 564 702, width=0.48\textwidth]{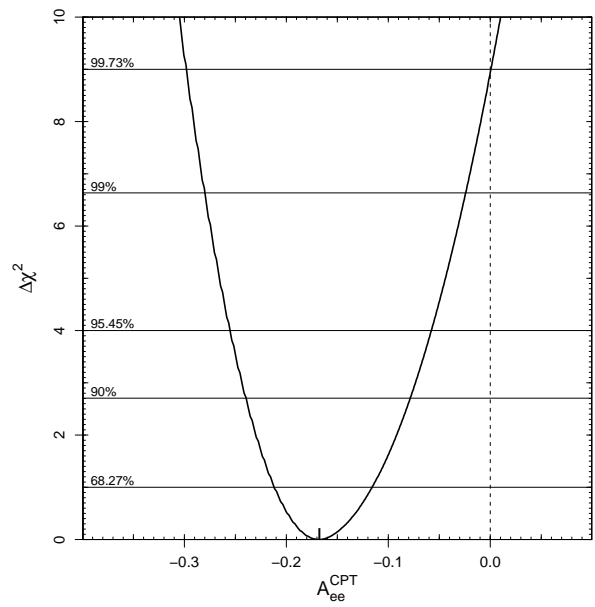}
\caption{ \label{chi-aee}
Marginal $\Delta\chi^2$
as a function of the CPT asymmetry
$A_{ee}^{\text{CPT}}$ in Eq.~(\ref{asy})
obtained from the fit of
MiniBooNE, Gallium and reactor data.
}
\end{center}
\end{figure}

In this section we consider a violation
\cite{hep-th/0012060,hep-ph/0108199,hep-ph/0204077,hep-ph/0308300,hep-ph/0309025,hep-th/0403037,hep-ph/0403285,hep-ph/0606154,hep-ph/0608007}
of the CPT equality
$ P_{\nu_{e}\to\nu_{e}} = P_{\bar\nu_{e}\to\bar\nu_{e}} $
as a possible explanation of the tension between
MiniBooNE and Gallium neutrino data on one side
and reactor antineutrino data on the other side
under the hypothesis of $\nu_{e}$ disappearance.
We quantify the amount of CPT violation through the asymmetry
\begin{equation}
A_{ee}^{\text{CPT}} \equiv P_{\nu_{e}\to\nu_{e}} - P_{\bar\nu_{e}\to\bar\nu_{e}}
\,.
\label{asy}
\end{equation}
Taking into account for completeness also the MiniBooNE neutrino data,
we minimized the least-squares function
$ \chi^2 = \chi^2_{\text{MB-$\nu$}} + \chi^2_{\text{Ga}} + \chi^2_{\text{Re}} + \chi^2_{\text{MB-$\bar\nu$}} $,
with $ P_{\nu_{e}\to\nu_{e}} $ replaced by $ P_{\bar\nu_{e}\to\bar\nu_{e}} $
in Eqs.~(\ref{chi-rea}) and (\ref{teo-anu}).
We obtained, for 26 degrees of freedom,
\begin{equation}
\chi^2_{\text{min}} = 39.9 
\,,
\qquad
\text{GoF} = 4.0\% 
\,,
\qquad
A_{ee}^{\text{CPT,bf}} = -0.17 
\,.
\label{asy-min}
\end{equation}
The relatively low goodness of fit is due to the relatively low goodness of fit of the
MiniBooNE neutrino and antineutrino data
(see Tabs.~\ref{tab-bef-neu} and \ref{tab-bef-anu}).
Figure~\ref{chi-aee} shows the marginal $\Delta\chi^2=\chi^2-\chi^2_{\text{min}}$
as a function of $A_{ee}^{\text{CPT}}$.
One can see that there is indication of CPT violation ($A_{ee}^{\text{CPT}}<0$) at 99.7\% CL.
The allowed intervals for $A_{ee}^{\text{CPT}}$ at
90\%,
95.45\%,
99\%, and
99.73\% CL
are, respectively,
\begin{equation}
A_{ee}^{\text{CPT}}
=
[ -0.24 , -0.08 ]
\,,\,
[ -0.25 , -0.06 ]
\,,\,
[ -0.28 , -0.02 ]
\,,\,
[ -0.30 , 0.00 ]
\,.
\label{asy-ranges}
\end{equation}
Let us emphasize that the possibility of CPT violation is extremely interesting
and should be explored in future experiments by measuring the CPT asymmetries
(see Ref.~\cite{Giunti-Kim-2007})
\begin{equation}
A_{\alpha\beta}^{\text{CPT}} \equiv P_{\nu_{\alpha}\to\nu_{\beta}} - P_{\bar\nu_{\beta}\to\bar\nu_{\alpha}}
\,,
\label{asy-all}
\end{equation}
with $\alpha,\beta=e,\mu$.
The realization of a CPT violation would have dramatic consequences for our understanding of
fundamental physics
\cite{hep-ph/0203261,hep-ph/0309309}.
Since the indication of CPT violation that we have found has been obtained under the hypothesis
of $\nu_{e}$ disappearance into sterile neutrinos,
it could be due to very exotic CPT-violating properties of the sterile neutrinos.

\section{Conclusions}
\label{022}
\nopagebreak

We have shown that the MiniBooNE low-energy neutrino anomaly
\cite{0704.1500,0812.2243}
can be explained by a VSBL $\nu_{e}$ disappearance
\cite{0707.4593}
which is compatible with that inferred from the Gallium radioactive source experiment anomaly
\cite{Anselmann:1995ar,Hampel:1998fc,Abdurashitov:1996dp,hep-ph/9803418,nucl-ex/0512041,0901.2200,Laveder:2007zz,hep-ph/0610352,0707.4593}.
There is a tension between this result and the absence of any indication of $\bar\nu_{e}$ disappearance in reactor neutrino data
\cite{Zacek:1986cu,Declais:1995su,hep-ex/0301017}
which could be due to statistical fluctuations, or to an underestimation of systematic uncertainties,
or to the inexistence of VSBL $\nu_{e}$ disappearance, or to a violation of the CPT equality
$ P_{\nu_{e}\to\nu_{e}} = P_{\bar\nu_{e}\to\bar\nu_{e}} $.
Considering the first possibility,
we have shown that the combined fit of
MiniBooNE neutrino, Gallium and reactor data
indicate that
$ P_{\nu_{e}\to\nu_{e}} < 1 $
at about $2\sigma$.

We have considered the
first MiniBooNE results on $\bar\nu_{\mu}\to\bar\nu_{e}$ oscillations \cite{Karagiorgi-FNAL-2008,0904.1958}.
Since the MiniBooNE antineutrino statistics is about five times smaller than the neutrino statistics,
the antineutrino data do not allow us to infer strong constraints on the VSBL $\nu_{e}$ disappearance hypothesis.
We have shown that MiniBooNE antineutrino data are compatible with the MiniBooNE neutrino data
and that adding the MiniBooNE antineutrino data to the combined fit of
MiniBooNE, Gallium and reactor data does not change the indication of
$ P_{\nu_{e}\to\nu_{e}} < 1 $
at about $2\sigma$.

We think that the possible VSBL $\nu_{e}$ disappearance discussed in this paper
is extremely interesting, because such disappearance is due to a $\Delta{m}^2$ which is much larger than
the well measured solar and atmospheric $\Delta{m}^2$'s
(see Ref.~\cite{0812.3161}).
Hence, our result indicate the possible existence of a light sterile neutrino
which is well beyond the Standard Model and would have
important consequences in
physics
(see Refs.~\cite{hep-ph/0609177,hep-ph/0611178,hep-ex/0701004,0704.0388,0705.0107,0706.1462,0707.2481,0710.2985}),
astrophysics
(see Refs.~\cite{0706.0399,0709.1937,0710.5180,0712.1816,0805.4014,0806.3029}),
and cosmology
(see Refs.~\cite{0711.2450,0810.5133,0812.2249}).

We have considered also the possibility of CPT violation as the source of the tension between the neutrino and antineutrino data.
For the asymmetry of the $\nu_{e}$ and $\bar\nu_{e}$ survival probabilities
we have obtained the best-fit value
$A_{ee}^{\text{CPT,bf}} = -0.17 $
and the constraint
$A_{ee}^{\text{CPT}}<0$ at 99.7\% CL.
Since the violation of CPT would have dramatic consequences four our understanding of fundamental physics,
we think that it should be investigated without prejudice in future experiments.
It could be due to very exotic CPT-violating properties of the sterile neutrinos into which
$\nu_{e}$ disappear in our hypothesis for the explanation of the MiniBooNE and Gallium anomalies.

Starting from 2010, at the same $L/E$ of MiniBoone,
the magnetic near detector at 280 m of the
T2K experiment at Tokai \cite{0810.2220}
will count $\nu_{e}$ events
with expected higher statistics
and similar $\nu_\mu$ background contamination.
A better $\nu_\mu$ background rejection will be possible
using the liquid Argon technology, as planned in the
the MicroBooNE proposal at Fermilab \cite{MicroBooNE-2008},
that will check the low-energy MiniBooNE anomaly.

The hypothesis of VSBL $\nu_{e}$ disappearance can be tested
with high accuracy by future experiments with pure electron neutrino beams.
The SAGE collaboration is planning to perform a new source experiment \cite{0901.2200}
in order to check the Gallium anomaly in Eq.~(\ref{ga}).
A similar measurement could be made in the LENS detector
\cite{Grieb:2006mp}.
Beta-beam experiments \cite{Zucchelli:2002sa,hep-ph/0602032}
with pure $\nu_{e}$ or $\bar\nu_{e}$ beams from nuclear decay
(see the reviews in Refs.~\cite{physics/0411123,hep-ph/0605033})
could investigate the disappearance of electron neutrinos and antineutrinos,
testing the CPT symmetry.
The same can be done, with higher precision, in
neutrino factory experiments
in which the beam is composed of
$\nu_{e}$ and $\bar\nu_{\mu}$,
from $\mu^{+}$ decay,
or
$\bar\nu_{e}$ and $\nu_{\mu}$,
from $\mu^{-}$ decay
(see the reviews in Refs.~\cite{hep-ph/0210192,physics/0411123}).
A Mossbauer neutrino experiment \cite{hep-ph/0601079},
with a $\bar\nu_{e}$ beam
produced in recoilless $^{3}\text{H}$ decay
and detected in recoilless $^{3}\text{He}$ antineutrino capture,
would be especially suited to investigate the VSBL disappearance of electron antineutrinos.

\section*{Acknowledgments}
\nopagebreak
We would like to thank
William C. Louis III for information about MiniBooNE neutrino data
and 
Carlo Bemporad and Donato Nicol\'o for information about the Chooz experiment.
We are grateful to Milla Baldo Ceolin for discussions and suggestions.
C. Giunti would like to thank the Department of Theoretical Physics of the University of Torino
for hospitality and support.

\raggedright


\begin{thebibliography}{10}

\bibitem{0812.2243}
MiniBooNE, A.A. Aguilar-Arevalo,
Phys. Rev. Lett. 102 (2009) 101802,
\href{http://arxiv.org/abs/0812.2243}{\url{arXiv:0812.2243}}.

\bibitem{0704.1500}
MiniBooNE, A. Aguilar-Arevalo et~al.,
Phys. Rev. Lett. 98 (2007) 231801,
\href{http://arxiv.org/abs/0704.1500}{\url{arXiv:0704.1500}}.

\bibitem{hep-ex/0104049}
LSND, A. Aguilar et~al.,
Phys. Rev. D64 (2001) 112007,
\href{http://arxiv.org/abs/hep-ex/0104049}{\url{arXiv:hep-ex/0104049}}.

\bibitem{Karagiorgi-FNAL-2008}
MiniBooNE, G. Karagiorgi,
(2008),
FNAL Seminar, 11 December 2008. URL:
\href{{http://theory.fnal.gov/jetp/talks/karagiorgi.pdf}}{\url{{http://theor%
y.fnal.gov/jetp/talks/karagiorgi.pdf}}}.

\bibitem{0904.1958}
MiniBooNE, A.A. Aguilar-Arevalo et~al.,
(2009), \href{http://arxiv.org/abs/0904.1958}{\url{arXiv:0904.1958}}.

\bibitem{0707.4593}
C. Giunti and M. Laveder,
Phys. Rev. D77 (2008) 093002,
\href{http://arxiv.org/abs/0707.4593}{\url{arXiv:0707.4593}}.

\bibitem{0708.1281}
J.A. Harvey, C.T. Hill and R.J. Hill,
Phys. Rev. Lett. 99 (2007) 261601,
\href{http://arxiv.org/abs/0708.1281}{\url{arXiv:0708.1281}}.

\bibitem{0712.1230}
J.A. Harvey, C.T. Hill and R.J. Hill,
Phys. Rev. D77 (2008) 085017,
\href{http://arxiv.org/abs/0712.1230}{\url{arXiv:0712.1230}}.

\bibitem{0709.4004}
A. Bodek,
(2007), \href{http://arxiv.org/abs/0709.4004}{\url{arXiv:0709.4004}}.

\bibitem{0711.1363}
A.E. Nelson and J. Walsh,
Phys. Rev. D77 (2008) 033001,
\href{http://arxiv.org/abs/0711.1363}{\url{arXiv:0711.1363}}.

\bibitem{Laveder:2007zz}
M. Laveder,
Nucl. Phys. Proc. Suppl. 168 (2007) 344,
NOW 2006.

\bibitem{hep-ph/0610352}
C. Giunti and M. Laveder,
Mod. Phys. Lett. A22 (2007) 2499,
\href{http://arxiv.org/abs/hep-ph/0610352}{\url{arXiv:hep-ph/0610352}}.

\bibitem{0901.2200}
SAGE, .J.N. Abdurashitov et~al.,
(2009), \href{http://arxiv.org/abs/0901.2200}{\url{arXiv:0901.2200}}.

\bibitem{Anselmann:1995ar}
GALLEX, P. Anselmann et~al.,
Phys. Lett. B342 (1995) 440.

\bibitem{Hampel:1998fc}
GALLEX, W. Hampel et~al.,
Phys. Lett. B420 (1998) 114.

\bibitem{Abdurashitov:1996dp}
SAGE, J.N. Abdurashitov et~al.,
Phys. Rev. Lett. 77 (1996) 4708.

\bibitem{hep-ph/9803418}
SAGE, J.N. Abdurashitov et~al.,
Phys. Rev. C59 (1999) 2246,
\href{http://arxiv.org/abs/hep-ph/9803418}{\url{arXiv:hep-ph/9803418}}.

\bibitem{nucl-ex/0512041}
J.N. Abdurashitov et~al.,
Phys. Rev. C73 (2006) 045805,
\href{http://arxiv.org/abs/nucl-ex/0512041}{\url{arXiv:nucl-ex/0512041}}.

\bibitem{0711.4222}
M.A. Acero, C. Giunti and M. Laveder,
Phys. Rev. D78 (2008) 073009,
\href{http://arxiv.org/abs/0711.4222}{\url{arXiv:0711.4222}}.

\bibitem{Giunti-Laveder-IP-09}
C. Giunti and M. Laveder,
(2009),
In preparation.

\bibitem{Giunti-Kim-2007}
C. Giunti and C.W. Kim,
{{Fundamentals of Neutrino Physics and Astrophysics}} (Oxford
University Press, 2007).

\bibitem{physics/0609129}
S.E. Kopp,
Phys. Rept. 439 (2007) 101,
\href{http://arxiv.org/abs/physics/0609129}{\url{arXiv:physics/0609129}},
NuFact Summer School.

\bibitem{0806.1449}
MiniBooNE, A.A. Aguilar-Arevalo et~al.,
(2008), \href{http://arxiv.org/abs/0806.1449}{\url{arXiv:0806.1449}}.

\bibitem{0706.0926}
MiniBooNE, A.A. Aguilar-Arevalo et~al.,
Phys. Rev. Lett. 100 (2008) 032301,
\href{http://arxiv.org/abs/0706.0926}{\url{arXiv:0706.0926}}.

\bibitem{Louis-private-09}
W.C. Louis,
(2009),
Private communication.

\bibitem{hep-ph/0304176}
M. Maltoni and T. Schwetz,
Phys. Rev. D68 (2003) 033020,
\href{http://arxiv.org/abs/hep-ph/0304176}{\url{arXiv:hep-ph/0304176}}.

\bibitem{0805.2098}
Y. Farzan, T. Schwetz and A.Y. Smirnov,
JHEP 07 (2008) 067,
\href{http://arxiv.org/abs/0805.2098}{\url{arXiv:0805.2098}}.

\bibitem{hep-ph/9710491}
J.N. Bahcall,
Phys. Rev. C56 (1997) 3391,
\href{http://arxiv.org/abs/hep-ph/9710491}{\url{arXiv:hep-ph/9710491}}.

\bibitem{hep-ph/0605186}
G. Fogli et~al.,
(2006),
\href{http://arxiv.org/abs/hep-ph/0605186}{\url{arXiv:hep-ph/0605186}},
3rd International Workshop on NO-VE: Neutrino Oscillations in Venice:
50 Years after the Neutrino Experimental Discovery, Venice, Italy, 7--10 Feb
2006.

\bibitem{nucl-th/9503017}
N. Hata and W. Haxton,
Phys. Lett. B353 (1995) 422,
\href{http://arxiv.org/abs/nucl-th/9503017}{\url{arXiv:nucl-th/9503017}}.

\bibitem{nucl-th/9804011}
W.C. Haxton,
Phys. Lett. B431 (1998) 110,
\href{http://arxiv.org/abs/nucl-th/9804011}{\url{arXiv:nucl-th/9804011}}.

\bibitem{Declais:1995su}
Bugey, B. Achkar et~al.,
Nucl. Phys. B434 (1995) 503.

\bibitem{Zacek:1986cu}
CalTech-SIN-TUM, G. Zacek et~al.,
Phys. Rev. D34 (1986) 2621.

\bibitem{hep-ex/0301017}
CHOOZ, M. Apollonio et~al.,
Eur. Phys. J. C27 (2003) 331,
\href{http://arxiv.org/abs/hep-ex/0301017}{\url{arXiv:hep-ex/0301017}}.

\bibitem{hep-th/0012060}
V.A. Kostelecky and R. Lehnert,
Phys. Rev. D63 (2001) 065008,
\href{http://arxiv.org/abs/hep-th/0012060}{\url{arXiv:hep-th/0012060}}.

\bibitem{hep-ph/0108199}
G. Barenboim et~al.,
JHEP 10 (2002) 001,
\href{http://arxiv.org/abs/hep-ph/0108199}{\url{arXiv:hep-ph/0108199}}.

\bibitem{hep-ph/0204077}
A. De~Gouvea,
Phys. Rev. D66 (2002) 076005,
\href{http://arxiv.org/abs/hep-ph/0204077}{\url{arXiv:hep-ph/0204077}}.

\bibitem{hep-ph/0308300}
A. Kostelecky and M. Mewes,
Phys. Rev. D70 (2004) 031902,
\href{http://arxiv.org/abs/hep-ph/0308300}{\url{arXiv:hep-ph/0308300}}.

\bibitem{hep-ph/0309025}
V.A. Kostelecky and M. Mewes,
Phys. Rev. D69 (2004) 016005,
\href{http://arxiv.org/abs/hep-ph/0309025}{\url{arXiv:hep-ph/0309025}}.

\bibitem{hep-th/0403037}
F.R. Klinkhamer and G.E. Volovik,
Int. J. Mod. Phys. A20 (2005) 2795,
\href{http://arxiv.org/abs/hep-th/0403037}{\url{arXiv:hep-th/0403037}}.

\bibitem{hep-ph/0403285}
F. Klinkhamer,
Jetp Lett. 79 (2004) 451,
\href{http://arxiv.org/abs/hep-ph/0403285}{\url{arXiv:hep-ph/0403285}}.

\bibitem{hep-ph/0606154}
T. Katori, V.A. Kostelecky and R. Tayloe,
Phys. Rev. D74 (2006) 105009,
\href{http://arxiv.org/abs/hep-ph/0606154}{\url{arXiv:hep-ph/0606154}}.

\bibitem{hep-ph/0608007}
P. Arias et~al.,
Phys. Lett. B650 (2007) 401,
\href{http://arxiv.org/abs/hep-ph/0608007}{\url{arXiv:hep-ph/0608007}}.

\bibitem{hep-ph/0203261}
G. Barenboim et~al.,
Phys. Lett. B537 (2002) 227,
\href{http://arxiv.org/abs/hep-ph/0203261}{\url{arXiv:hep-ph/0203261}}.

\bibitem{hep-ph/0309309}
O. Greenberg,
Found. Phys. 36 (2006) 1535,
\href{http://arxiv.org/abs/hep-ph/0309309}{\url{arXiv:hep-ph/0309309}}.

\bibitem{0812.3161}
M. Maltoni and T. Schwetz,
(2008), \href{http://arxiv.org/abs/0812.3161}{\url{arXiv:0812.3161}},
IDM2008, Aug. 18-22, 2008, Stockholm, Sweden.

\bibitem{hep-ph/0609177}
G. Karagiorgi et~al.,
Phys. Rev. D75 (2007) 013011,
\href{http://arxiv.org/abs/hep-ph/0609177}{\url{arXiv:hep-ph/0609177}}.

\bibitem{hep-ph/0611178}
C. Grieb, J. Link and R.S. Raghavan,
In Phys. Rev. \cite{Grieb:2006mp}, p. 093006,
\href{http://arxiv.org/abs/hep-ph/0611178}{\url{arXiv:hep-ph/0611178}}.

\bibitem{hep-ex/0701004}
D.C. Latimer, J. Escamilla and D.J. Ernst,
Phys. Rev. C75 (2007) 042501,
\href{http://arxiv.org/abs/hep-ex/0701004}{\url{arXiv:hep-ex/0701004}}.

\bibitem{0704.0388}
A. Donini et~al.,
JHEP 12 (2007) 013,
\href{http://arxiv.org/abs/0704.0388}{\url{arXiv:0704.0388}}.

\bibitem{0705.0107}
M. Maltoni and T. Schwetz,
Phys. Rev. D76 (2007) 093005,
\href{http://arxiv.org/abs/0705.0107}{\url{arXiv:0705.0107}}.

\bibitem{0706.1462}
S. Goswami and W. Rodejohann,
JHEP 10 (2007) 073,
\href{http://arxiv.org/abs/0706.1462}{\url{arXiv:0706.1462}}.

\bibitem{0707.2481}
A. Bandyopadhyay and S. Choubey,
(2007), \href{http://arxiv.org/abs/0707.2481}{\url{arXiv:0707.2481}}.

\bibitem{0710.2985}
T. Schwetz,
JHEP 02 (2007) 011,
\href{http://arxiv.org/abs/0710.2985}{\url{arXiv:0710.2985}}.

\bibitem{0706.0399}
R.L. Awasthi and S. Choubey,
Phys. Rev. D76 (2007) 113002,
\href{http://arxiv.org/abs/0706.0399}{\url{arXiv:0706.0399}}.

\bibitem{0709.1937}
S. Choubey,
JHEP 12 (2007) 014,
\href{http://arxiv.org/abs/0709.1937}{\url{arXiv:0709.1937}}.

\bibitem{0710.5180}
D. Boyanovsky, H.J. de~Vega and N. Sanchez,
Phys. Rev. D77 (2008) 043518,
\href{http://arxiv.org/abs/0710.5180}{\url{arXiv:0710.5180}}.

\bibitem{0712.1816}
G. Gentile, H.S. Zhao and B. Famaey,
(2007), \href{http://arxiv.org/abs/0712.1816}{\url{arXiv:0712.1816}}.

\bibitem{0805.4014}
G.W. Angus,
(2008), \href{http://arxiv.org/abs/0805.4014}{\url{arXiv:0805.4014}}.

\bibitem{0806.3029}
A. Donini and O. Yasuda,
(2008), \href{http://arxiv.org/abs/0806.3029}{\url{arXiv:0806.3029}}.

\bibitem{0711.2450}
O. Civitarese and M.E. Mosquera,
Phys. Rev. C77 (2008) 045806,
\href{http://arxiv.org/abs/0711.2450}{\url{arXiv:0711.2450}}.

\bibitem{0810.5133}
A. Melchiorri et~al.,
JCAP 0901 (2008) 036,
\href{http://arxiv.org/abs/0810.5133}{\url{arXiv:0810.5133}}.

\bibitem{0812.2249}
M.A. Acero and J. Lesgourgues,
Phys. Rev. D79 (2009) 045026,
\href{http://arxiv.org/abs/0812.2249}{\url{arXiv:0812.2249}}.

\bibitem{0810.2220}
T2K, T. Lindner et~al.,
(2008), \href{http://arxiv.org/abs/0810.2220}{\url{arXiv:0810.2220}},
ICHEP08.

\bibitem{MicroBooNE-2008}
MicroBooNE, H. Chen et~al.,
(2008),
URL:
\href{{http://www-microboone.fnal.gov/Documents/MicroBooNE_addendum/MicroBoo%
NEAddendum_030308.pdf}}{\url{{http://www-microboone.fnal.gov/Documents/MicroBo%
oNE_addendum/MicroBooNEAddendum_030308.pdf}}}.

\bibitem{Grieb:2006mp}
C. Grieb, J. Link and R.S. Raghavan,
Phys. Rev. D75 (2007) 093006,
\href{http://arxiv.org/abs/hep-ph/0611178}{\url{arXiv:hep-ph/0611178}}.

\bibitem{Zucchelli:2002sa}
P. Zucchelli,
Phys. Lett. B532 (2002) 166.

\bibitem{hep-ph/0602032}
C. Rubbia et~al.,
Nucl. Instrum. Meth. A568 (2006) 475,
\href{http://arxiv.org/abs/hep-ph/0602032}{\url{arXiv:hep-ph/0602032}}.

\bibitem{physics/0411123}
Neutrino Factory/Muon Collider, C. Albright et~al.,
(2004),
\href{http://arxiv.org/abs/physics/0411123}{\url{arXiv:physics/0411123}}.

\bibitem{hep-ph/0605033}
C. Volpe,
J. Phys. G34 (2007) R1,
\href{http://arxiv.org/abs/hep-ph/0605033}{\url{arXiv:hep-ph/0605033}}.

\bibitem{hep-ph/0210192}
M. Apollonio et~al.,
(2002),
\href{http://arxiv.org/abs/hep-ph/0210192}{\url{arXiv:hep-ph/0210192}}.

\bibitem{hep-ph/0601079}
R.S. Raghavan,
(2006),
\href{http://arxiv.org/abs/hep-ph/0601079}{\url{arXiv:hep-ph/0601079}}.

\end{thebibliography}


\clearpage

\begin{table}[ht!]
\begin{center}
\begin{tabular}{cccccc}
$j$
&
Energy Range [MeV]
&
$N^{\text{cal}}_{\nu_{e},j}$
&
$N^{\text{cal}}_{\nu_{\mu},j}$
&
$N^{\text{cal}}_{\nu,j}$
&
$N^{\text{exp}}_{\nu,j}$
\\
\hline
1 & $ 200 - 300 $ & $ 18.8 $ & $ 168.0 $ & $ 186.8 $ & $ 232 $ \\
2 & $ 300 - 375 $ & $ 23.4 $ & $ 85.0 $ & $ 108.4 $ & $ 156 $ \\
3 & $ 375 - 475 $ & $ 40.6 $ & $ 79.8 $ & $ 120.4 $ & $ 156 $ \\
4 & $ 475 - 550 $ & $ 31.6 $ & $ 32.7 $ & $ 64.3 $ & $ 79 $ \\
5 & $ 550 - 675 $ & $ 50.8 $ & $ 39.7 $ & $ 90.5 $ & $ 81 $ \\
6 & $ 675 - 800 $ & $ 49.9 $ & $ 17.9 $ & $ 67.8 $ & $ 70 $ \\
7 & $ 800 - 950 $ & $ 52.7 $ & $ 17.8 $ & $ 70.5 $ & $ 63 $ \\
8 & $ 950 - 1100 $ & $ 44.3 $ & $ 13.3 $ & $ 57.6 $ & $ 65 $ \\
9 & $ 1100 - 1300 $ & $ 42.5 $ & $ 9.9 $ & $ 52.4 $ & $ 62 $ \\
10 & $ 1300 - 1500 $ & $ 33.8 $ & $ 5.3 $ & $ 39.1 $ & $ 34 $ \\
11 & $ 1500 - 3000 $ & $ 56.7 $ & $ 15.5 $ & $ 72.2 $ & $ 71 $ \\
\hline
\end{tabular}
\caption{ \label{tab-neu-data}
MiniBooNE neutrino data \protect\cite{0812.2243,Louis-private-09}.
The six columns give:
1) bin number;
2) reconstructed neutrino energy range;
3) number of expected $\nu_{e}$-induced events;
4) number of expected misidentified $\nu_{\mu}$-induced events;
5) total number of expected events;
6) measured number of events.
}
\end{center}
\end{table}

\begin{table}[ht!]
\begin{center}
\begin{tabular}{cccccc}
&
&
MB-$\nu$
&
MB-$\nu$+Ga
&
Re
&
MB-$\nu$+Ga+Re
\\
\hline
 & $\chi^{2}_{\text{min}}$ &27.2 &34.0 &2.9 &36.9 \\
No Osc. & NDF &10 &11 &7 &18 \\
 & GoF &0.2\% &0.04\% &89.8\% &0.5\% \\
 & $f_{\nu}^{\text{bf}}$ &1.15 &1.15 & &1.15 \\
\hline & $\chi^{2}_{\text{min}}$ &17.7 &20.1 &2.9 &31.7 \\
 & NDF &9 &10 &6 &17 \\
Osc. & GoF &3.8\% &2.8\% &82.7\% &1.7\% \\
 & $P_{\nu_{e}\to\nu_{e}}^{\text{bf}}$ &0.72 &0.83 &1.0 &0.93 \\
 & $f_{\nu}^{\text{bf}}$ &1.31 &1.24 & &1.19 \\
\hline & $\Delta\chi^{2}_{\text{min}}$ & &2.4 & &11.1 \\
PG & NDF & &1 & &2 \\
 & GoF & &12.4\% & &0.4\% \\
\hline
\end{tabular}
\caption{ \label{tab-bef-neu}
Values of
$\chi^{2}$,
number of degrees of freedom (NDF) and
goodness-of-fit (GoF)
for the fit of different combinations of
MiniBooNE neutrino (MB-$\nu$), Gallium (Ga) and
reactor (Re) data.
The first four lines correspond to the case of no oscillations (No Osc.).
The following five lines correspond to the case of oscillations (Osc.).
The last three lines give the parameter goodness-of-fit (PG) \protect\cite{hep-ph/0304176}.
}
\end{center}
\end{table}

\begin{table}[ht!]
\begin{center}
\setlength{\tabcolsep}{5pt}
\begin{tabular}{lcccccc}
& BF
& 68.27\%
& 90\%
& 95.45\%
& 99\%
& 99.73\%
\\
\hline
MB-$\nu$ & 0.72 & 0.65 - 0.80 & 0.60 - 0.86 & 0.58 - 0.89 & 0.54 - 0.95 & 0.52 - 0.99 \\
MB-$\nu$+Ga & 0.83 & 0.79 - 0.88 & 0.76 - 0.91 & 0.75 - 0.92 & 0.72 - 0.95 & 0.70 - 0.97 \\
Re & 1.00 & 0.97 - 1.00 & 0.94 - 1.00 & 0.93 - 1.00 & 0.91 - 1.00 & 0.89 - 1.00 \\
MB-$\nu$+Ga+Re & 0.93 & 0.90 - 0.96 & 0.89 - 0.98 & 0.88 - 0.99 & 0.86 - 1.00 & 0.85 - 1.00 \\
MB-$\bar\nu$ & 1.00 & 0.83 - 1.00 & 0.70 - 1.00 & 0.63 - 1.00 & 0.54 - 1.00 & 0.47 - 1.00 \\
MB-$\bar\nu$+Re & 1.00 & 0.97 - 1.00 & 0.95 - 1.00 & 0.93 - 1.00 & 0.91 - 1.00 & 0.89 - 1.00 \\
MB & 0.76 & 0.69 - 0.84 & 0.65 - 0.90 & 0.62 - 0.93 & 0.59 - 0.98 & 0.56 - 1.00 \\
MB+Ga+Re & 0.93 & 0.91 - 0.96 & 0.89 - 0.98 & 0.88 - 0.99 & 0.86 - 1.00 & 0.85 - 1.00 \\
\hline
\end{tabular}
\caption{ \label{tab-pee}
Best-fit values (BF) and allowed ranges of
$P_{\nu_{e}\to\nu_{e}}$
at the indicated value of confidence level
for the fits in Tabs.~\ref{tab-bef-neu} and \ref{tab-bef-anu}.
}
\end{center}
\end{table}

\begin{table}[ht!]
\begin{center}
\begin{tabular}{cccccc}
$j$
&
Energy Range [MeV]
&
$N^{\text{cal}}_{\bar\nu_{e},j}$
&
$N^{\text{cal}}_{\bar\nu_{\mu},j}$
&
$N^{\text{cal}}_{\bar\nu,j}$
&
$N^{\text{exp}}_{\bar\nu,j}$
\\
\hline
1 & $ 200 - 300 $ & $ 4.3 $ & $ 22.5 $ & $ 26.8 $ & $ 24 $ \\
2 & $ 300 - 375 $ & $ 4.2 $ & $ 11.4 $ & $ 15.6 $ & $ 21 $ \\
3 & $ 375 - 475 $ & $ 7.0 $ & $ 11.1 $ & $ 18.1 $ & $ 16 $ \\
4 & $ 475 - 550 $ & $ 5.4 $ & $ 4.7 $ & $ 10.1 $ & $ 14 $ \\
5 & $ 550 - 675 $ & $ 6.9 $ & $ 5.0 $ & $ 11.8 $ & $ 22 $ \\
6 & $ 675 - 800 $ & $ 7.4 $ & $ 3.0 $ & $ 10.3 $ & $ 9 $ \\
7 & $ 800 - 950 $ & $ 7.7 $ & $ 3.0 $ & $ 10.7 $ & $ 5 $ \\
8 & $ 950 - 1100 $ & $ 6.3 $ & $ 2.7 $ & $ 9.0 $ & $ 7 $ \\
9 & $ 1100 - 1300 $ & $ 5.8 $ & $ 2.3 $ & $ 8.1 $ & $ 5 $ \\
10 & $ 1300 - 1500 $ & $ 4.2 $ & $ 1.6 $ & $ 5.8 $ & $ 7 $ \\
11 & $ 1500 - 3000 $ & $ 9.5 $ & $ 2.5 $ & $ 12.0 $ & $ 14 $ \\
\hline
\end{tabular}
\caption{ \label{tab-anu-data}
MiniBooNE antineutrino data extracted from the figure in page 55 of Ref.~\protect\cite{Karagiorgi-FNAL-2008}.
The six columns give:
1) bin number;
2) reconstructed antineutrino energy range;
3) number of expected $\bar\nu_{e}$-induced events;
4) number of expected misidentified $\bar\nu_{\mu}$-induced events;
5) total number of expected events;
6) measured number of events.
}
\end{center}
\end{table}

\begin{table}[ht!]
\begin{center}
\begin{tabular}{cccccc}
&
&
MB-$\bar\nu$
&
MB-$\bar\nu$+Re
&
MB
&
MB+Ga+Re
\\
\hline
 & $\chi^{2}_{\text{min}}$ &16.9 &19.8 &44.1 &53.8 \\
No Osc. & NDF &10 &17 &21 &29 \\
 & GoF &7.6\% &28.5\% &0.2\% &0.3\% \\
 & $f_{\bar\nu}^{\text{bf}}$ &1.08 &1.08 &1.08 &1.08 \\
\hline & $\chi^{2}_{\text{min}}$ &16.9 &19.8 &36.7 &48.9 \\
 & NDF &9 &16 &19 &27 \\
Osc. & GoF &5.0\% &23.0\% &0.9\% &0.6\% \\
 & $P_{\nu_{e}\to\nu_{e}}^{\text{bf}}$ &1.00 &1.00 &0.76 &0.93 \\
 & $f_{\bar\nu}^{\text{bf}}$ &1.08 &1.08 &1.19 &1.10 \\
 & $f_{\nu}^{\text{bf}}$ & & &1.28 &1.19 \\
\hline & $\Delta\chi^{2}_{\text{min}}$ & &0.0 &2.1 &8.3 \\
PG & NDF & &1 &1 &3 \\
 & GoF & &100.0\% &14.8\% &4.1\% \\
\hline
\end{tabular}
\caption{ \label{tab-bef-anu}
Values of
$\chi^{2}$,
number of degrees of freedom (NDF) and
goodness-of-fit (GoF)
for the fit of
MiniBooNE antineutrino (MB-$\bar\nu$),
MiniBooNE antineutrino and reactor (MB-$\bar\nu$+Re),
MiniBooNE neutrino and antineutrino (MB)
and
MiniBooNE neutrino and antineutrino, Gallium and reactor (MB+Ga+Re)
data.
The first four lines correspond to the case of no oscillations (No Osc.).
The following six lines correspond to the case of oscillations (Osc.).
The last three lines give the parameter goodness-of-fit (PG) \protect\cite{hep-ph/0304176}.
}
\end{center}
\end{table}


\clearpage

\end{document}